# Hybrid Centralized-Distributed Resource Allocation Based on Deep Reinforcement Learning for Cooperative D2D Communications

Yang Yu[1]* and XIAOQING TANG[2]
[1]Electronic Information School, Hubei Three Gorges Polytechnic, Yichang 443000, China
[2]School of Artificial Intelligence, Hubei University, Wuhan 430062, China

**ABSTRACT** Device-to-device (D2D) technology enables direct communication between adjacent devices within cellular networks. Due to its high data rate, low latency, and performance improvement in spectrum and energy efficiency, it has been widely investigated and applied as a critical technology in 5G New Radio (NR). In addition to conventional overlay and underlay D2D communications, cooperative D2D communication, which can achieve a win-win situation between cellular users (CUs) and D2D users (DUs) through cooperative relaying technique, has attracted extensive attention from academic and industrial circles in the past decade. This paper delves into optimizing joint spectrum allocation, power control, and link-matching between multiple CUs and DUs for cooperative D2D communications, using weighted sum energy efficiency (WSEE) as the performance metric to address the challenges of green communication and sustainable development. This integer programming problem can be decomposed into a classic weighted bipartite graph matching and a series of nonconvex spectrum allocation and power control problems between potentially matched cellular and D2D link pairs. To address this issue, we propose a hybrid centralized-distributed scheme based on deep reinforcement learning (DRL) and the Kuhn-Munkres (KM) algorithm. Leveraging the latter, the CUs and DUs autonomously optimize spectrum allocation and power control by only utilizing local information. Then, the base station (BS) determines the link matching. Simulation results reveal that it achieves near-optimal performance and significantly enhances the network convergence speed with low signaling overheads. In addition, we also propose and utilize cooperative link sets for corresponding D2D links to accelerate the proposed scheme and reduce signaling exchange further.

## I. INTRODUCTION

### A. BACKGROUND

With tremendous advances in wireless communication technology, mobile traffic has grown unprecedentedly over the past decades, bringing an urgent need to improve spectrum efficiency for 5G and beyond cellular networks. Reducing cell radius, which leads to small cells, is a conventional solution to tighten the frequency reuse factor. In this way, the proximity of a base station (BS) and its corresponding cellular users (CUs) can achieve a higher data rate with lower latency and power consumption. However, to support small cells, the requirements of additional infrastructures will result in considerable costs in construction and maintenance [1]. As an alternative, device-to-device (D2D) technology enables adjacent devices to communicate directly, effectively alleviating BSs' burden through traffic offloading, which makes it a promising paradigm for improving spectrum efficiency (SE) at a low cost [2].

The idea of cognitive radio has profoundly motivated D2D communications [3]. As secondary users (SUs) do in cognitive radio networks, D2D users (DUs) typically utilize CUs' spectrum in two approaches of spectrum sharing: overlay and underlay [4]. In the overlay approach, BSs reserve a portion of cellular frequency bands that DUs can access exclusively. Thus, no interference exists between cellular links and D2D links. Compared with the overlay approach, the underlay approach allows a CU and DU to transmit simultaneously on the same band, inevitably causing mutual interference [5]. Whether using the overlay or underlay approach, it can be seen that the authorized CUs' benefit suffers from a decrease in available cellular bands or interference from neighboring DUs. In order to compensate for CUs, cooperative D2D communications have been proposed [6]. Leveraging the cooperative relaying technique, DUs can play as relays, helping CUs with poor channel conditions finish their data transmission first and then access the corresponding CUs' cellular bands exclusively. In this process, the DUs exchange their partial energy budget for the usage rights of the cellular bands, and the CUs guarantee their required quality of service (QoS), which achieves a win-win situation between CUs and DUs [6]–[8]. Therefore, in this paper, to balance the benefits of CUs and DUs, we focus on wireless resource allocation in cooperative D2D communications [8] but do not consider the switching between the overlay, underlay, and cooperative relay modes as the authors did in [6].

### B. RELATED WORK

Resource allocation for cooperative D2D communication networks has been extensively researched. Various network-centric and user-centric schemes [9] under mathematical frameworks such as convex optimization [7], [10], contract theory [11], [12], and game theory [8] have been extensively proposed. Specifically, the authors of [7] investigated cooperative D2D overlaying communications for public-interest and self-interest DUs with spectrum-power trading to maximize DUs' weighted sum energy efficiency (WSEE). In [10], the authors proposed non-orthogonal multiple access (NOMA) assisted schemes integrating simultaneous wireless information and power transfer (SWIPT) to optimize the overall system throughput. [11] investigated the contract design under incomplete information. In [12], the authors further extended the work of [11] to the field of cognitive Internet of Things (IoT) networks using orthogonal frequency-division multiple access (OFDMA). The authors in [8] proposed a two-timescale scheme in order to improve spectrum efficiency (SE) through matching game theory.

In recent years, cellular networks have evolved to be ultra-dense and heterogeneous [13], involving intelligent mobile devices that adaptively adjust their transmission parameters according to surrounding dynamic wireless environments [14]. Faced with this trend, designing resource allocation algorithms needs to balance signaling overhead and convergence speed. However, most existing schemes require relatively high costs for signaling exchanges. For example, frequent interaction exists between user equipments (UEs) until the convergence of the schemes based on game theory, leading to excessive overheads [15], [16]. Moreover, network-centric schemes entail the exchange of massively instantaneous signals, such as global channel state information (CSI) [17]. Thus, many existing schemes are typically not suitable for real-time operation.

Deep reinforcement learning (DRL) has evolved in the past decade as an effective and promising approach to handling decision-making problems. Since DRL can reveal hidden environmental patterns via continuously interacting with uncertain and dynamic environments [18], it is suitable for dealing with the signaling overhead issue in cellular networks. Its recently increasing applications for efficient wireless resource management [19]-[21] motivate us to adopt it in cooperative D2D communication networks.

Traditional reinforcement learning (RL) technologies, e.g., Q-learning, can handle problems with small action-state spaces. However, as cellular and D2D links increase, the action-state space will grow synchronously. When the link number is large enough, these methods will no longer apply to large-scale networks. Fortunately, DRL surmounts this shortcoming with the aid of neural networks (NNs) [22], [23]. So far, there have already been some successes in channel selection [24], [25] and power allocation [26], [27]. Specifically, a dynamic multi-channel access problem was considered in [24], and DRL was used to maximize the expectation of long-term successful transmission numbers without exact information on system dynamics. The authors in [25] investigated the communications between industrial devices with different priorities. A multi-channel access scheme based on DRL was proposed to guarantee that high-priority devices have the highest channel access probability while minimizing queuing delay. In [26], the authors demonstrated the potential of DRL on transmit power control in large-scale networks and developed a model-free DRL-based distributively dynamic scheme. The scheme is especially suitable for scenarios where channel state information (CSI) delay cannot be ignored. [27] proposed a multi-agent DRL framework for dynamic power control to maximize downlink network throughput. The proposed algorithms outperformed the state-of-art model-based methods.

Moreover, due to its ability to withstand uncertain and dynamic environments, DRL has also been gradually used in fields such as cooperative spectrum sensing [28], computation offloading [29], and UAV-Assisted Communications [30].

### C. PERFORMANCE METRIC: WSEE

Due to economic, environmental, and health concerns, sustainable development has become crucial in designing and operating present and future wireless communication networks. Energy efficiency (EE), the ratio between global achievable throughput and power consumption, is a frequently used metric for measuring system cost-effectiveness. In this paper, balancing the EEs between different nodes according to their priority levels is also one of our goals. However, taking system EE as the objective function ignores the priorities of different nodes in heterogeneous networks, which cannot meet this requirement. As an alternative, WSEE considers individual links' EEs while maintaining global performance. With these considerations, this paper adopts WSEE as the objective function.

Maximizing system EE produces a single-ratio fractional programming (FP) problem [31], which conventional methods can handle under a mature framework. In contrast, WSEE maximization belongs to multiple-ratio FP problems. Although recent literature has proposed some solution frameworks [32], [33], they are often hard to solve and require global CSIs. Fortunately, the emergence of DRL greatly facilitates WSEE optimization, as mentioned above.

*D. CONTRIBUTIONS AND ORGANIZATIONS*

This paper focuses on a D2D communication network where multiple D2D and cellular links work in cooperative relay mode without co-channel interference. A problem of joint spectrum allocation, power control, and link-matching is formulated to maximize system WSEE. Unfortunately, this problem is challenging due to its non-convexity and NP-hardness. Applying the state-of-the-art method, FP, to it is more complex than the well-investigated power control problems [32] owing to the coupling of spectrum allocation, power control, and link matching. In particular, link matching, which involves integer programming, will further increase the complexity of its solution. Furthermore, a scheme based on FP must be deployed centrally, which requires instantaneous global CSI, leading to excessive signaling overheads and poor extensibility.

Although DRL-based methods have been extensively investigated separately in channel selection [24], [25] or power control [20], [26], [27], a scheme of joint spectrum allocation, power control, and link matching for cooperative D2D communications has yet to be considered. In addition, since cooperative D2D communications involve a large number of CSIs, a central controller, such as the BS, acting as the RL agent, will bring a significant amount of signaling overheads [20]. On the other hand, although distributed schemes [26], [27] employing multi-agents can alleviate this issue to a certain extent, they also need to improve the frequent information exchange between agents and slow convergence speed. Thus, developing an efficient scheme based on DRL to achieve extensibility and reduce signaling overheads is an urgent issue.

To solve this, we analyze the cooperative D2D communication process and notice that this integer programming problem can be broken down into a series of nonconvex spectrum allocation and power control problems and a weighted bipartite graph matching problem between potentially matched cellular and D2D link pairs. On this basis, we develop a hybrid centralized-distributed scheme based on DRL and the Kuhn-Munkres (KM) algorithm. Moreover, cooperative link sets have also been applied to improve our proposed scheme.

Thus, the main contributions of our paper can be summarized below:

1. To the best of our knowledge, it is the first time that DRL has been introduced to handle resource allocation in classical three-phase cooperative D2D communications.

2. Under the proposed hybrid centralized-distributed scheme, each agent' computational overheads will be independent of the network size. Leveraging DRL, a D2D transmitter (DT), which acts as an RL agent, can operate autonomously and independently to optimize spectrum allocation and power control between itself and its adjacent cellular links. Then, the BS determines the link matching according to the utility function matrix reported by the agents. Eventually, the proposed scheme will reduce signaling overheads significantly compared with the existing network-centric algorithms and can achieve near-optimal performance.

3. We introduce the concept of cooperative link sets for the agents. With them, the proposed scheme can be further accelerated with fewer signaling overheads.

The rest of our paper is organized as follows: In section 2, we present the system model. Then, a joint problem of spectrum allocation, power control, and link-matching is formulated to maximize system WSEE. In section 3, we develop a hybrid centralized-distributed scheme based on the KM algorithm and DRL to solve the problem. Cooperative link sets are also utilized to accelerate the proposed scheme and reduce signaling overhead. In section 4, we present simulation results to evaluate the performance of our proposed scheme. Finally, we conclude this paper in section 5.

## II. SYSTEM MODEL AND PROBLEM FORMULATION

*A. COOPERATIVE SPECTRUM SHARING MODEL BETWEEN CELLULAR AND D2D LINKS*

Since a BS has relatively high transmit power and a sufficient energy budget, cellular downlink transmission hardly requires performing cooperative D2D communication. This paper focuses on the uplink cellular transmission due to the limited energy budget for mobile terminals. As depicted in Fig. 1, a single-cell scenario with a BS and multiple CUs is considered. Several D2D pairs reuse the cellular wireless frequency resource to implement their communication. One D2D pair consists of a DT and a D2D receiver (DR). Moreover, we assume the transmission direction of each DT remains unchanged for simplicity. Referring to a CU-BS uplink as a cellular link and a D2D pair as a D2D link, we denote $\mathcal{C} = \{c_1, ..., c_M\}$ and $\mathcal{D} = \{d_1, ..., d_N\}$ as the sets of the cellular links and D2D links in this system, where $M = |\mathcal{C}|$ and $N = |\mathcal{D}|$. For the convenience of derivation, we define $\mathcal{M} = \{1, ..., M\}$ and $\mathcal{N} = \{1, ..., N\}$, respectively. Some main symbols in this paper are included in Table I.

In 5G New Radio (NR), resource blocks (RBs) are the basic units of wireless time-frequency resources allocated to UEs. Generally, the BS allocates a single RB or contiguous RBs to one CU. In this paper, referring to [6] and [20], we suppose that the number of RBs equals $M$, and the BS only assigns one RB to one cellular link. Furthermore, each D2D link is allowed to access only one RB, while each cellular link also allows only one D2D link to occupy its RB, so the orthogonality of wireless resources can be preserved. For simplicity, we assume that the cellular and D2D links can only operate in cooperative relay mode. For example, this situation occurs when the cellular links suffer from poor channel conditions and can not complete direct data transmission; meanwhile, the D2D links can only access idle authorized spectrum. Throughout the rest of this paper, we adopt the decoding and forwarding (DF) protocol and take the case that the DTs act as relays to describe the process of cooperative spectrum sharing.

The frame structure adopted by the considered system is illustrated in Fig. 2. Specifically, each frame starts with the preprocessing stage and follows with the cooperative spectrum-sharing subframes. During the preprocessing stage, pilot transmission, CSI reporting, and resource allocation are carried out sequentially. The CUs and the DTs first send pilots [34]. Then, the BS and the DRs estimate the channel gains and report CSI, respectively. With the CSI, the resource allocation can be determined. The details of cooperative spectrum sharing between one matched cellular and D2D link pair are described as follows.

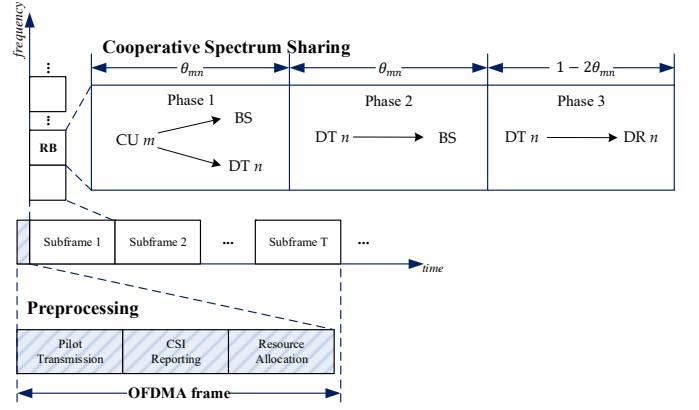

FIGURE 2. The frame structure for the considered system.

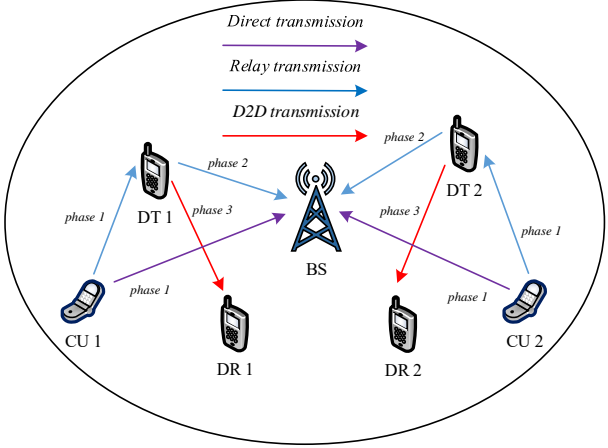

FIGURE 1. An illustration of the system model with two CUs and two D2D pairs.

TABLE I
SOME IMPORTANT NOMENCLATURE

| Symbol | Definition |
| --- | --- |
| $\mathcal{C}, \mathcal{D}$ | Sets of cellular and D2D links |
| $M, N$ | Number of cellular and D2D links |
| $c_m$ | Cellular link $m$ |
| $d_n$ | D2D link $n$ |
| $g_{mn}$ | Channel gain between CU $m$ and DT $n$ |
| $g_{mb}$ | Channel gain between CU $m$ and the BS |
| $g_{nb}$ | Channel gain between DT $n$ and the BS |
| $g_{nn}$ | Channel gain between DT $n$ and DR $n$ |
| $p_{c,mn}$ | Transmit power of CU $m$ (if $c_m$ and $d_n$ are matched) |
| $p_{r,mn}$ | Relaying power of DT $n$ |
| $p_{d,mn}$ | Transmit power of DT $n$ for its own data |
| $n_0$ | Additive white Gaussian noise power density |
| $\theta_{mn}$ | Spectrum-sharing factor in the time domain |
| $\alpha_{mn}$ | Matching indicator for $c_m$ and $d_n$ |
| $Q_c, Q_d$ | Minimum achievable SE required for cellular and D2D links |
| $\mu_m, \nu_n$ | Constant weights for $c_m$ and $d_n$ |
| $p_{min}, p_{max}$ | Minimum and maximum transmit power |
| $\Delta p, \Delta \theta$ | Granularities of power and spectrum-sharing factor levels |
| $u_{mn}$ | WSEE of the matched cellular-D2D link pair $c_m$ and $d_n$ |
| $\mathcal{E}_n$ | Cooperative link set of $d_n$ |
| $r_{n1}, r_{n2}$ | Cooperative ranges of $d_n$ for DT $n$ and the BS |

When cellular link $c_m$ and D2D link $d_n$ are matched, their cooperative spectrum-sharing subframes are divided into three phases. In phase one, CU $m$ broadcasts to the BS and DT $n$ using power $p_{c,mn}$. In the second phase, DT $n$ decodes and forwards the information received in phase one to the BS using power $p_{r,mn}$. In phase three, owing to the help from DT $n$, CU $m$ has completed its uplink transmission. DT $n$ gets an opportunity to fulfill its data transmission to its corresponding DR using power $p_{d,mn}$ on the RB of CU $m$. Since we adopt the DF protocol, the duration of phase one equals that of phase two. Thus, we can denote the duration of these three phases as $\theta_{mn}, \theta_{mn}, 1 - 2\theta_{mn}$ of an RB in the time domain, and $\theta_{mn} \in (0, 0.5)$ is the spectrum-sharing factor.

According to [8], the achievable SEs of $c_m$ and $d_n$ are given as

$$SE_{c,mn} = \theta_{mn} log_2[1 + min(\gamma_{mn} p_{c,mn}, \gamma_{mb} p_{c,mn} + \gamma_{nb} p_{r,mn})], \quad (1)$$

$$SE_{d,mn} = (1 - 2\theta_{mn}) log_2(1 + \gamma_{nn} p_{d,mn}), \quad (2)$$

where $\gamma_{mn} = g_{mn}/n_0$, $\gamma_{mb} = g_{mb}/n_0$, $\gamma_{nb} = g_{nb}/n_0$, and $\gamma_{nn} = g_{nn}/n_0$. Moreover, $n_0$ is the noise power density; $g_{mn}$, $g_{mb}$, $g_{nb}$, and $g_{nn}$ denote the channel gains between CU $m$ and DT $n$, CU $m$ and the BS, DT $n$ and the BS, DT $n$ and DR $n$, respectively. Correspondingly, the transmit power consumption of $c_m$ and $d_n$ are expressed as

$$P_{c,mn} = \theta_{mn} p_{c,mn}, \quad (3)$$

$$P_{d,mn} = \theta_{mn} p_{r,mn} + (1 - 2\theta_{mn}) p_{d,mn}. \quad (4)$$

If $c_m$ and $d_n$ are not matched, based on previous assumptions, we have $SE_{c,mn} = SE_{d,mn} = 0$ and $P_{c,mn} = P_{d,mn} = 0$.

### B. PROBLEM FORMULATION FOR WSEE MAXIMIZATION

In recent years, green communication, which aims to minimize the energy consumption of network components while maintaining quality of service (QoS) requirements, has become an emerging strategy to reduce carbon footprint and quest for sustainability regarding human health and environmental conditions. Generally, EE is the most critical performance metric in line with green communication. System EE is the overall achievable SE ratio to power consumption. If we omit the circuit power consumption, the overall system EE is expressed as

$$EE = \frac{\sum_{m=1}^{M}\sum_{n=1}^{N}\alpha_{mn}(SE_{c,mn}+SE_{d,mn})}{\sum_{m=1}^{M}\sum_{n=1}^{N}\alpha_{mn}(P_{c,mn}+P_{d,mn})}, \quad (5)$$

where $\alpha_{mn} \in \{0,1\}$ indicates whether $c_m$ and $d_n$ are matched or not. If $\alpha_{mn} = 1$, $c_m$ matches $d_n$, otherwise, $\alpha_{mn} = 0$. Denoting $\boldsymbol{p}_c = \{p_{c,mn} | m \in \mathcal{M}, n \in \mathcal{N}\}$, $\boldsymbol{p}_r = \{p_{r,mn} | m \in \mathcal{M}, n \in \mathcal{N}\}$, and $\boldsymbol{p}_d = \{p_{d,mn} | m \in \mathcal{M}, n \in \mathcal{N}\}$ as the sets of the transmit power; $\boldsymbol{\alpha} = \{\alpha_{mn} | \alpha_{mn} = 0,1; m \in \mathcal{M}, n \in \mathcal{N}\}$ as the set of the matching indicators; $\boldsymbol{\theta} = \{\theta_{mn} | \theta_{mn} \in (0,0.5); m \in \mathcal{M}, n \in \mathcal{N}\}$ as the set of the spectrum-sharing factors, the optimization problem for EE maximization can be expressed as follows.

$$\max_{\{\boldsymbol{p}_c,\boldsymbol{p}_r,\boldsymbol{p}_d,\boldsymbol{\alpha},\boldsymbol{\theta}\}} EE \quad (6)$$

s.t. C1: $SE_{c,mn} \geq Q_c$,

C2: $SE_{d,mn} \geq Q_d$,

C3: $p_{min} \leq p_{c,mn} \leq p_{max}$,

C4: $p_{min} \leq p_{r,mn} \leq p_{max}$,

C5: $p_{min} \leq p_{d,mn} \leq p_{max}$,

C6: $\sum_{m=1}^{M}\sum_{n=1}^{N}\alpha_{mn} \in \{0,1\}$,

C7: $0 < \theta_{mn} < 0.5$,

where $m \in \mathcal{M}$, $n \in \mathcal{N}$; C1-C7 are the constraints the cellular and D2D links should obey during the cooperative spectrum-sharing subframes. Specifically, C1 and C2 guarantee the minimum achievable SE requirements of the cellular and D2D links; C3, C4, and C5 constrain the lower and upper limits of the transmit power, denoted by $p_{min}$ and $p_{max}$, respectively; C6 refers to the matching rule mentioned in Subsection II.A.

Solving (6) is challenging since this problem is NP-hard and belongs to mixed integer nonlinear programming (MINLP). The corresponding optimal solution can be found using an exhaustive search but with high computational complexity. Alternatively, a near-optimal solution via FP can be obtained, which has been widely proposed and acknowledged [31], [32]. Typically, a centralized resource allocation algorithm must be performed at the BS, which increases the computational load of the BS rapidly with the size of the network.

RL is a machine-learning technique used to solve decision-making problems, and it has found widespread applications in wireless communication networks. RL-based resource allocation schemes can adjust to the dynamic wireless environment using a trial-and-error method, which is less challenging than supervised learning-based schemes that require a collection of label data [34]. This consideration motivates us to develop an optimization scheme based on RL to achieve a near-optimal solution for this problem.

Nevertheless, besides the computational load issues of the BS mentioned earlier, some areas still need improvement in the centralized RL solution. For example, the authors in [20] proposed an RL-based algorithm with the system EE as the objective function. The BS acts as the learning agent and will deal with high-dimensional discrete state and action spaces if there are massive cellular or D2D links, which makes learning and acting intractable in time. It will entail a long preprocessing stage to observe all combinations of user states and train the agent. Furthermore, any changes in the number of cellular or D2D links, such as mobile handover, result in the alteration of state and action spaces, requiring the RL agent to be retrained at short intervals due to the changing structure of the objective function.

On the other hand, CUs usually have higher priority than DUs. However, adopting EE as the objective function does not reflect the priorities of different users. As an alternative to EE, WSEE considers both system-centric and fairness-centric EE. It is an effective metric for networks with heterogeneous nodes because network operators can utilize more degrees of freedom from predefined weights to customize different users' performance. More importantly, applying WSEE as the objective function can effectively avoid the issues related to high-dimensional, variable state space and action space that occurred in [20].

The individual link EE of either $c_m$ or $d_n$ involved in the cooperative spectrum sharing is defined as the ratio of achievable SE to power consumption, which is given by

$$EE_{c,mn} = \frac{SE_{c,mn}}{P_{c,mn}}, \quad (7)$$

$$EE_{d,mn} = \frac{SE_{d,mn}}{P_{d,mn}}. \quad (8)$$

Therefore, we can express the system WSEE as

$$WSEE = \sum_{m=1}^{M}\sum_{n=1}^{N}\alpha_{mn}(\mu_m EE_{c,mn} + \nu_n EE_{d,mn}), \quad (9)$$

where $\mu_m$ and $\nu_n$ are the positive constant weights of $c_m$ and $d_n$, predefined by the network operator determining the priorities of different links. By replacing the objective function of (6) with (9), we update the optimization problem as follows.

$$\max_{\{\boldsymbol{p}_c,\boldsymbol{p}_r,\boldsymbol{p}_d,\boldsymbol{\alpha},\boldsymbol{\theta}\}} WSEE \quad (10)$$

s.t. C1, C2, C3, C4, C5, C6, C7.

Since C6 means a one-to-one correspondence between the cellular links $\mathcal{C}$ and the D2D links $\mathcal{D}$, and noting the structure of the objective function in (10), we find that (10) can be decomposed into two levels. The lower level comprises a series of spectrum allocation and power control, while the higher level is a maximum-weighted bipartite matching. This decoupling, which avoids a MINLP problem, i.e., (6) and naturally leads to a hybrid centralized-distributed solution, is one critical reason we introduced WSEE as the optimization goal. Besides, in this way, we do not have to deal with the high-dimensional state and action spaces when utilizing DRL to solve this problem.

## III. HYBRID CENTRALIZED-DISTRIBUTED SCHEME FOR COOPERATIVE D2D COMMUNICATIONS

### A. DRL-based Spectrum Allocation and Power Control

Based on the above analysis, we first delve into the subproblem of spectrum allocation and power control during cooperative spectrum sharing. Our primary focus is on one matched cellular-D2D link pair. When $c_m$ cooperates with $d_n$ on its corresponding RB, the optimization problem to maximize their WSEE is expressed as follows.

$$\max_{\{p_{c,mn}, p_{r,mn}, p_{d,mn}, \theta_{mn}\}} \mu_m EE_{c,mn} + \nu_n EE_{d,mn} \quad (11)$$

s.t. C1, C2, C3, C4, C5, C7.

By adding a negative sign before $\mu_m EE_{c,mn} + \nu_n EE_{d,mn}$ and minimizing the new objective function, we can transform (11) into its standard form. Then, we can prove that this standard-form optimization problem is nonconvex as follows. Using the Karush-Kuhn-Tucker (KKT) [35] conditions to solve this problem is tricky.

**Proposition:** The standard-form optimization problem of (11) is nonconvex.

**Proof:** We define an auxiliary parameter $\beta = \gamma_{nn}$ and the following auxiliary function

$$f(x,y) = (2y - 1) \log_2(1 + \beta x). \quad (12)$$

The Hessian matrix of $f(x,y)$ is calculated as

$$\mathbf{H}(f) = \begin{bmatrix} \frac{\beta^2(1-2y)}{(1+\beta x)^2 \ln 2} & \frac{2\beta}{(1+\beta x)\ln 2} \\ \frac{2\beta}{(1+\beta x)\ln 2} & 0 \end{bmatrix}. \quad (13)$$

We can obtain its eigenvalues by solving the characteristic equation of $\mathbf{H}$, i.e., $\det(\lambda \mathbf{I} - \mathbf{H}) = 0$, where $\mathbf{I}$ is an identity matrix, as follows.

$$\lambda_1 = \frac{\beta^2(1-2y)}{2(1+\beta x)^2 \ln 2} + \sqrt{\left[\frac{\beta^2(1-2y)}{2(1+\beta x)^2 \ln 2}\right]^2 + \left[\frac{2\beta}{(1+\beta x)\ln 2}\right]^2}, \quad (14)$$

$$\lambda_2 = \frac{\beta^2(1-2y)}{2(1+\beta x)^2 \ln 2} - \sqrt{\left[\frac{\beta^2(1-2y)}{2(1+\beta x)^2 \ln 2}\right]^2 + \left[\frac{2\beta}{(1+\beta x)\ln 2}\right]^2}. \quad (15)$$

It is easy to see $\lambda_1 > 0$ and $\lambda_2 < 0$ over $\{(x,y) | x \in \mathbf{R}_+, y \in (0, 0.5)\}$. $\mathbf{H}$ is not positive semi-definite, so function $f(x,y)$ is not convex [35]. Since $Q_d - SE_{d,mn} = Q_d + f(p_{d,mn}, \theta_{mn})$, C2 is nonconvex constraint. Therefore, the standard-form optimization problem of (11) is nonconvex.
**Q.E.D.**

In fact, (11) belongs to the family of multiple-ratio FP problems, which are generally hard to handle and require global CSIs [32], [33]. On the other hand, it has been proven that achieving high system performance through a static resource allocation strategy in a dynamic cellular network is next to impossible. Comparatively, real-time learning is more suitable for future wireless networks. It enables intelligent agents to make decisions through interactions with their surrounding environments, requiring no prior knowledge.

Resource allocation can be implemented through a central controller, e.g., the BS, using centralized schemes or individual UEs using distributed schemes. RL-based approaches are extensively utilized to deal with resource allocation problems in various domains in a centralized and distributed manner. While centralized schemes perform well for small networks, their performance degrades as the number of UEs increases due to massive signaling overheads from CSI feedback. In contrast, by distributing computations to each UE, distributed schemes maintain the same individual computational overheads as the network scales up. These considerations reassure us that distributed schemes are preferred, especially for ultra-dense networks. An ideal distributed framework allows each D2D link to autonomously decide its resource allocation, thereby generating a scalable scheme.

Observing that only four variables are involved in (11), a scalable distributed scheme will naturally yield if we restrict the decision-making to each cellular-D2D link pair. Meanwhile, the issue of the high-dimensional discrete state space and action space occurring in [20] can be avoided under our system model, where the wireless resources are allocated in an orthogonal manner. Hence, we map our system to a multi-agent DRL framework, where DTs are considered learning agents interacting with the environment, in contrast to the case where the BS is employed as a single learning agent.

However, to maximize (11) under a distributed framework, in theory, if each D2D link selects its matched link from all the cellular links in the network, it will require plenty of network information exchange with high individual computational cost, which conflicts with an efficient resource allocation strategy. Therefore, after we have derived the proposed resource allocation scheme, we will constrain D2D link $d_n$ to select its matched link within a set, hereafter termed the cooperative link set $\mathcal{E}_n$. Later, we will discuss its generation criteria.

On the other hand, due to the orthogonality of spectrum resources in our system, conflicts may arise between D2D links when they independently select their matched cellular links from each other. Such conflicts can be prevented to a certain extent through interaction and negotiation among multiple agents. However, it will introduce extra communication overheads and lead to multiple iterations among agents. To avoid this, we propose a hybrid resource allocation scheme combining the characteristics of centralized and distributed manners. In our scheme, each DT periodically broadcasts its action-reward values, and the BS decides the link matching, which means different agents have no interaction, unlike in typical multi-agent schemes. During the learning process, the DTs are assumed to operate in a mode such as 3GPP NR V2X Mode 2 [36]. The framework of our proposed multi-agent DRL-based scheme will be introduced thereinafter.

1) DRL COMPONENTS

As stated earlier, adopting the proposed hybrid scheme avoids cooperative decision-making among multiple agents. To further simplify the derivation process, this paper assumes that all agents have access to the exact estimation of CSIs in each decision-making instant. This subsection models the spectrum allocation and power control subproblem (11) as an interactive decision-making process and defines the DRL components.

**Agent:** Agent $mn$ implements the spectrum allocation and power control between $c_m$ and $d_n$. We assume that DT $n$ acts as agent $mn$ in the proposed scheme. The components of DRL corresponding to agent $mn$ include $(s_{mn,t}; a_{mn,t}; r_{mn,t}; s_{mn,t+1})$, which means agent $mn$ in state $s_{mn,t}$ selects action $a_{mn,t}$ based on a specific policy at step $t$. After that, it receives reward $r_{mn,t}$ and moves to state $s_{mn,t+1}$ at training step $t+1$.

**State:** State space $\mathcal{S}_{mn}$ should reflect signal-to-noise ratio (SNR) at DT $n$, DR $n$, the BS, current transmit power and spectrum-sharing factor. We assume that the information about the SNR at DR $n$, the BS, and the transmit power of CU $m$ are broadcast on the common control channel (CCCH) so the agent $mn$ can acquire them in real time. On the other hand, according to (1)-(2), we notice that the SNR $\gamma_{mn}p_{c,mn}$, $\gamma_{mb}p_{c,mn}$, $\gamma_{nb}p_{r,mn}$, and $\gamma_{nn}p_{d,mn}$ are linearly dependent on the channel gains. Therefore, we formally define $\mathcal{S}_{mn} = \{g_{mn}, g_{mb}, g_{nb}, g_{nn}, p_{c,mn}, p_{r,mn}, p_{d,mn}, \theta_{mn}\}$.

**Action and Policy:** Action space $\mathcal{A}_{mn}$ denotes the possible transmit power and spectrum-sharing factor adjustment operation. This paper considers discrete power levels widely applied in existing communication standards, e.g., LTE-A and 5G. Meanwhile, applying discrete power levels is beneficial if a transmitter has limited capabilities owing to hardware constraints, such as machine-type communications [37]. Generally speaking, optimization problems with discrete power levels are more challenging than those involving continuous power levels, as the former will result in integer programming. Fortunately, we can efficiently address this challenge within the DRL framework. Similarly, we assume the spectrum-sharing factor levels are also discrete.

We define the action $a_{mn,t} = \{p_{c,mn}, p_{r,mn}, p_{d,mn}, \theta_{mn}\} \in \mathcal{A}_{mn}$ at each $t$. $p_{c,mn}$, $p_{r,mn}$, $p_{d,mn}$ are chosen from the possible power levels, namely, $\{p_{min}, p_{min}\Delta p, ..., p_{max}\}$, where $\Delta p = \sqrt[I-1]{p_{max}/p_{min}}$ and $I$ denotes the number of power levels. Similarly, $\theta_{mn}$ selects its value from the possible spectrum-sharing factor levels, namely, $\{\Delta\theta, 2\Delta\theta, ..., 0.5 - \Delta\theta\}$, where $\Delta\theta = 0.5/(L+1)$ and $L$ express the numbers of spectrum-sharing factor levels.

To balance exploration and exploitation, the agent selects actions following the $\epsilon$-greedy algorithm [18].

**Reward:** One solution for an RL problem with constraints is to transform it into a constraint Markov Decision Process (CMDP) and solve it through primal-dual optimization (PDO) or constraint policy optimization (CPO) [38], [39]. Another approach is incorporating domain knowledge to modify reward functions, i.e., reward shaping. In this paper, we adopt the latter approach.

If all the constraints can be met, the RL process will encourage higher WSEE. Therefore, we naturally consider defining the reward function as the WSEE of the link pair. However, if any constraint of problem (11) is unsatisfied, this action results in a penalty. Since the state space $\mathcal{S}_{mn}$ has already implied that C3, C4, C5, and C7 can be adhered to, we only need to consider C1 and C2. We assume the violations of C1 or C2 are tolerable during the training of the RL agent but should try our best to guarantee C1 and C2. Considering the above factors, we can design the reward function with two predefined positive constants, $\varphi$ and $\phi$.

$$r_{mn} = (\mu_m EE_{c,mn} + \nu_n EE_{d,mn})\left[1 + \varphi \frac{min(SE_{c,mn} - Q_c, 0)}{Q_c} + \phi \frac{min(SE_{d,mn} - Q_d, 0)}{Q_d}\right]. \quad (16)$$

For the sake of clarity, we omit the subscript $mn$ when we introduce the DRL framework throughout the rest of this article.

According to the Bellman equation [18], the optimal strategy is to choose an action that maximizes the optimal action-value function

$$Q^*(s, a) = \mathbb{E}\left[r_{t+1} + \gamma \max_{a'} Q^*(s_{t+1}, a') | s_t = s, a_t = a\right]. \quad (17)$$

where $\gamma$ denotes the discount rate.

Since the iterations are discrete, acquiring the accurate value is directly impractical. A NN is applied as an approximator to evaluate the action-value function

$$Q(s_t, a_t; \boldsymbol{W}) \approx Q^*(s, a). \quad (18)$$

2) DQN TRAINING

The agent regards state $\mathcal{S}$ as its input and generates the evaluated Q-value for the state-action pair. A deep Q network (DQN) is invoked to evaluate the Q-value table and represent WSEE. Agent $mn$ at DT $n$ trains the NNs, as shown in Fig. 3. We store the training data $e_t = (s_t; a_t; r_t; s_{t+1})$ in replay memory $\mathfrak{D}$, where $r_t$ is calculated using (16). During the training process, a minibatch $e_i = (s_i; a_i; r_i; s_{i+1})$ is sampled from $\mathfrak{D}$.

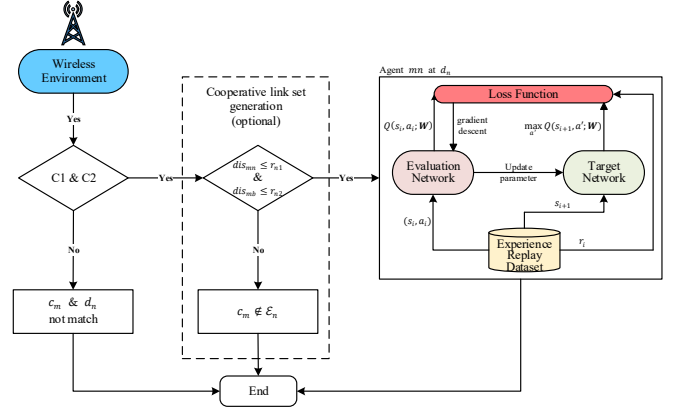

**FIGURE 3. The proposed DQN enabled spectrum allocation and power control.**

The DQN is composed of two NNs, namely, the evaluation network and the target network. The former inputs state $s_i$ and action $a_i$, generating the evaluated Q-value. Correspondingly, the latter regards $s_{i+1}$ as the input and outputs the target Q-value for minibatch $i$. We define the loss function as follows to minimize the difference between the estimated and target Q-values.

$$Loss(\mathbf{W}) = \mathbb{E}\left[\left(q_{target} - Q(s_i, a_i; \mathbf{W})\right)^2\right], \quad (19)$$

where $q_{target} = r_i + \gamma \max_{a'} Q(s_{i+1}, a'; \mathbf{W}^*)$ denotes the target Q-value. Moreover, $\mathbf{W}^*$ and $\mathbf{W}$ denote target and evaluation network weights. We optimize $\mathbf{W}$ using the gradient descent method [18]. The details of the DQN training process are displayed in Table II.

TABLE II
DQN TRAINING FOR AN INDIVIDUAL AGENT

1: **Begin**
2:   **for** each training episode, **do**
3:     **for** each training step, **do**
4:       Select action $a_t$ from action space $\mathcal{A}$ based on the $\epsilon$-greedy algorithm;
5:       Implement $a_t$, calculate reward $r_t$ using (16) and then observe $s_{t+1}$;
6:       Store $(s_t; a_t; r_t; s_{t+1})$ into $\mathfrak{D}$;
7:       Replay memory:
8:       Sample a random minibatch $(s_i; a_i; r_i; s_{i+1})$;
9:       Calculate $q_{target}$;
10:      Carry out a gradient descent step on $[q_{target} - Q(s_i, a_i; \mathbf{W})]^2$;
11:     **end**
12:     Update $\mathbf{W}^*$ by $\mathbf{W}^* = \mathbf{W}$;
13: **end**

### B. LINK MATCHING

The BS determines the matching between the cellular links $\mathcal{C}$ and D2D links $\mathcal{D}$ in our proposed scheme. Due to the matching rule C6, we can formulate the link matching, which aims to maximize the system WSEE, as a classical maximum bipartite matching with the weights $\{u_{mn} = \mu_m EE_{c,mn} + \nu_n EE_{d,mn} | m \in \mathcal{M}, n \in \mathcal{N}\}$ obtained from (11). According to all the agents' feedback, a weight matrix can be formed at the BS as

$$\mathbf{U} = \begin{bmatrix} u_{11} & u_{12} & \cdots & u_{1N} \\ u_{21} & u_{22} & \cdots & u_{2N} \\ \vdots & \vdots & \ddots & \vdots \\ u_{M1} & u_{M2} & \cdots & u_{MN} \end{bmatrix}. \quad (20)$$

In general, when we have obtained matrix $\mathbf{U}$, the optimal matching strategy $\boldsymbol{\alpha}^*$ between $\mathcal{C}$ and $\mathcal{D}$ can be reached through the classical KM algorithm with a time complexity of $O(\max\{M, N\}^3)$ [40]. Some other matching algorithms with lower computational complexity are suitable for specific scenarios, i.e., a two-dimensional matching proposed in [41] to determine the optimal spectrum resource allocation for relay-to-receiver links. However, it is not the point of this paper. We still apply the classical KM algorithm because of its generality. In future work, we will be committed to further enhancing the efficiency of the matching process in our proposed scheme.

### C. COOPERATIVE LINK SET

As discussed above, a D2D link selects its potential matched link from its cooperative link set, which should contain as few cellular links as possible to reduce signal exchange between the D2D link and its potential matched cellular links. At the same time, the performance loss when D2D link $d_n$ selects its matched cellular link from its cooperative link set $\mathcal{E}_n$ should be minimized compared to considering all the cellular links. These two principles above should be efficiently met in this paper. As reflected in (20), we should make the matrix $\mathbf{U}$ as sparse as possible while decreasing system performance as little as possible. To this end, we attempt to generate $\mathcal{E}_n$ for $d_n$ using a heuristic approach.

Firstly, we can quickly determine whether $c_m$ should be excluded from $\mathcal{E}_n$ by setting $p_{c,mn} = p_{r,mn} = p_{d,mn} = p_{max}$. According to C1 and C2, we can obtain

$$\frac{Q_c}{log_2[1+p_{max}min(g_{mn},g_{mb}+g_{nb})/n_0]} \leq \theta_{mn} \leq \frac{1}{2} - \frac{Q_d}{2\,log_2(1+p_{max}g_{nn}/n_0)}. \quad (21)$$

If (21) does not meet C7, the feasible domain of problem (11) is $\emptyset$.

Then, returning to (1) and (2), we notice that the performance of one matched cellular-D2D link pair, i.e., $c_m$ and $d_n$, is highly correlated with $g_{mn}$, $g_{mb}$, $g_{nb}$, and $g_{nn}$. Wherein $g_{mn}$ and $g_{mb}$ depend on the selected cellular link $c_m$. Hence, we naturally consider incorporating cellular link $c_m$ into $\mathcal{E}_n$ if $g_{mn}$ and $g_{mb}$ are above two predefined thresholds, namely $g_{n1}$ and $g_{n2}$, respectively. However, the reasonable values of $g_{n1}$ and $g_{n2}$ are challenging to determine. Channel gains generally comprise large-scale and small-scale components. Specifically, pass loss and shadowing belong to large-scale fading, which remains unchanged for a long time. This article mainly considers static and quasi-static channels dominated by large-scale fading. Thus, alternatively, we can approximately select potentially matched cellular link $c_m$ to form $\mathcal{E}_n$, as shown in Fig. 4, if the distances CU $m$ to DT $n$ and from CU $m$ to the BS are within two predefined cooperative ranges, i.e., $dis_{mn} \leq r_{n1}$ and $dis_{mb} \leq r_{n2}$, respectively. $r_{n1}$ and $r_{n2}$ should be functions of CU density, and the BS decides their values. In this process, each CU and each DT should operate in 3GPP NR V2X Mode 2 and periodically broadcast their current locations.

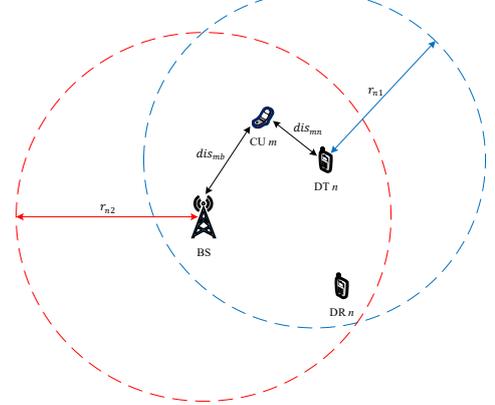

**FIGURE 4.** The criteria to select $c_m$ for $\mathcal{E}_n$.

## IV. SIMULATION RESULTS AND DISCUSSION
This section validates the performance of our proposed scheme by simulations. Some important simulation parameters are presented as follows. $n_0 = -174$ dBm/Hz, $p_{min} = -40$ dBm, $p_{max} = 23$ dBm, and $\mu_m = \nu_n = 1$. The granularities of power and spectrum-sharing factor levels are $\Delta p = 3$ dB and $\Delta \theta = 0.05$, respectively. The required minimum SEs are $Q_c = 5$ bps/Hz and $Q_d = 3$ bps/Hz. We present them and more simulation parameters in Table III. The channel gains are assumed to obey a stationary distribution. A simplified path loss model with a path loss exponent 3.8 is utilized. The exploration is set to be $\epsilon = 1 - 0.8\,t/500$, where $t$ denotes the current episode and the maximum number of training episodes is 500.

We first focus on a single-cell cellular network with a radius $R = 500$ m. The CUs, DTs, and DRs are randomly located within the cell. We use the Monte Carlo simulation based on 1000 runs. The cooperative ranges are both set as $r_{n1} = r_{n2} = 3R/4 = 375$ m for $\forall n \in \mathcal{N}$.

The proposed scheme is compared with three other schemes: the optimal scheme via an exhaustive search, random resource allocation, and the proposed scheme with the cooperative link sets. Specifically, the optimal scheme, which checks all possible combinations to obtain the optimal solution, is regarded as an upper-performance benchmark. The computational complexity of the optimal scheme will increase rapidly as the number of links increases, making it impossible to simulate the optimal solution if there are many cellular or D2D links. Therefore, we set $N = 5 \sim 15$ as an instance to show the performance gap between the optimal and proposed schemes. The random resource allocation randomly selects the spectrum-sharing factors and the transmit power. In addition, the cellular and D2D links are also matched randomly. This scheme is employed as a lower-performance benchmark and utilized in many works. DT $n$ selects its potential matched cellular links within the cooperative link sets in the last comparing scheme.

TABLE III
SIMULATION PARAMETERS

| Symbol | Value |
| --- | --- |
| Cell radius, $R$ | 500 m |
| Cooperative range, $r_{n1}$ | 375 m |
| Cooperative range, $r_{n2}$ | 375 m |
| Path loss exponent, $\alpha$ | 3.8 |
| Gaussian noise spectral density, $n_0$ | -174 dBm/Hz |
| Minimum transmit power, $p_{min}$ | -40 dBm |
| Maximum transmit power, $p_{max}$ | 23 dBm |
| Granularity of the power levels, $\Delta p$ | 3 dB |
| Granularity of the spectrum-sharing factor levels, $\Delta\theta$ | 0.05 |
| Number of the cellular links, $M$ | 10 |
| Number of the D2D links, $N$ | 5 ~ 15 |
| Minimum achievable SE required for the cellular links, $Q_c$ | 5 bps/Hz |
| Minimum achievable SE required for the D2D links, $Q_d$ | 3 bps/Hz |
| Weight for the cellular links, $\mu_m$ | 1 |
| Weight for the D2D links, $v_n$ | 1 |
| Constant corresponding to the cellular links in the reward function, $\varphi$ | 1 |
| Constant corresponding to the D2D links in the reward function, $\phi$ | 1 |

Fig. 5 shows a snapshot of node locations when $M = 10$ and $N = 5$. Fig. 6 examines the overall system WSEE versus the number of D2D links. The optimal scheme based on exhaustive search acquires the best performance. The random scheme performs the worst because spectrum allocation, power control, and link-matching are randomly configured but not jointly optimized. We can see that the WSEEs of all the schemes increase as the number of D2D links increases. In line with our expectations, the performance gap between the proposed and the optimal schemes is smaller than the gaps between the other considered schemes and the optimal scheme. It is evident that the proposed scheme achieves near-optimal performance, while the proposed scheme with the cooperative link sets has the second most minor performance loss compared with the optimal scheme. For example, when $N = 10$, the proposed scheme achieves more than 95% performance of the optimal scheme. It far outperforms the random scheme. Besides, the proposed scheme with the cooperative link sets achieves an average of about 85% WSEE of the optimal scheme and performs better when $N$ is small.

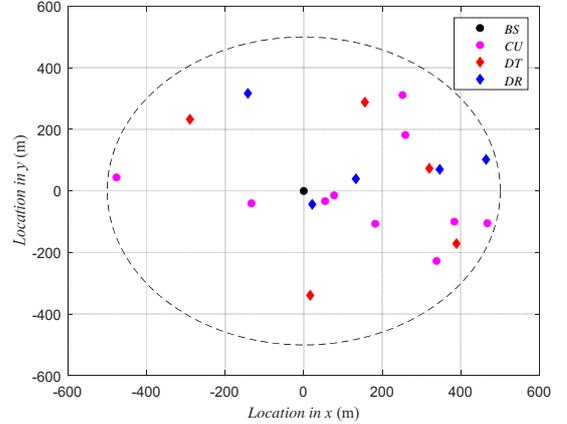

FIGURE 5. A snapshot of node locations when $M = 10$ and $N = 5$.

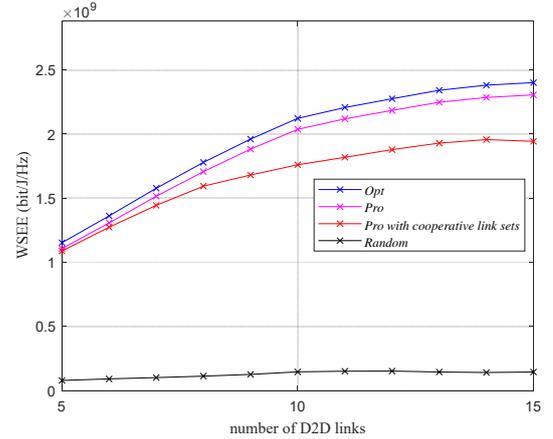

FIGURE 6. The system WSEE vs. the number of D2D links.

Fig. 7 shows the number of non-zero elements in matrix $\mathbf{U}$ versus the number of D2D links for the proposed schemes with and without the cooperative sets. We can observe from this figure that applying the cooperative sets can significantly reduce the number of non-zero elements, meaning fewer agents will be activated in the DRL process. At the same time, less information about action-reward values will be reported to the BS.

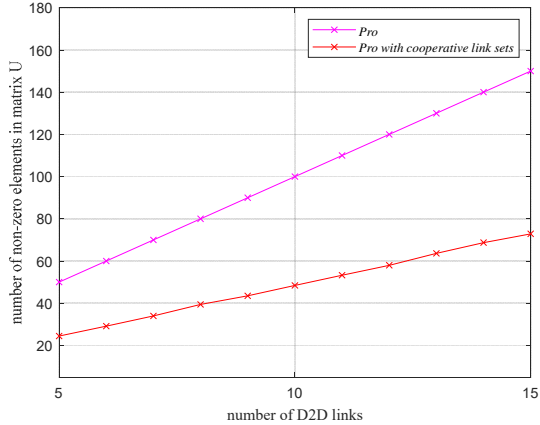

**FIGURE 7.** The number of non-zero elements in matrix U vs. the number of D2D links.

The above results, revealed in Figs. 6 and 7, demonstrate that introducing the cooperative link sets reduces signaling overheads and computational cost, while the performance of the proposed scheme can be well guaranteed by setting reasonable cooperative ranges.

Then, we focus on a single matched link pair and analyze the training process of one single DQN. It is worth noting that cooperative link sets are not employed by the proposed scheme in this case. We set the distances between the CU and the BS, the DT and the BS, and the DT and the DR as 1000 m, 500 m, and 500 m, respectively. Moreover, the distance between the CU and the DT varies from 500 to 1500 m. These settings simulate the situation when the D2D pair moves along a specific trajectory. Fig. 8 exhibits the WSEE versus the episode number $t$ at the initial distance between the CU and the DT. It increases quickly as the episode number increases at the beginning. After about 110 episodes, the WSEEs remain almost unchanged.

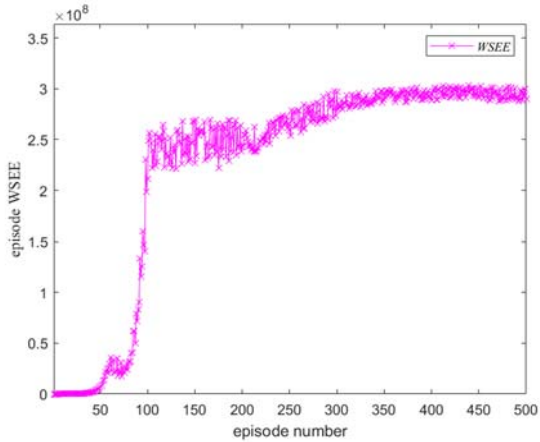

**FIGURE 8.** Episode WSEE vs. the episode number.

Finally, Fig. 9 compares the computation time of the considered schemes versus the distance between the CU and the DT. This simulation measures the computation time using a laptop with a Ryzen 9 7945HX CPU, a GeForce RTX 4080 GPU, and 64 GB memory. We notice that all the schemes have low computation time except for the optimal one. In stark contrast, the computation time of the optimal one is higher by several orders of magnitude than the proposed scheme, e.g., about 5.4 s, when the distance between the CU and the DT equals 500 m. Thus, the proposed scheme balances performance and computation time well. We also noticed that the convergence time of the proposed scheme is less when the distance between CU and DT is not the initial value, 500m. This phenomenon highlights the learning ability of the agents.

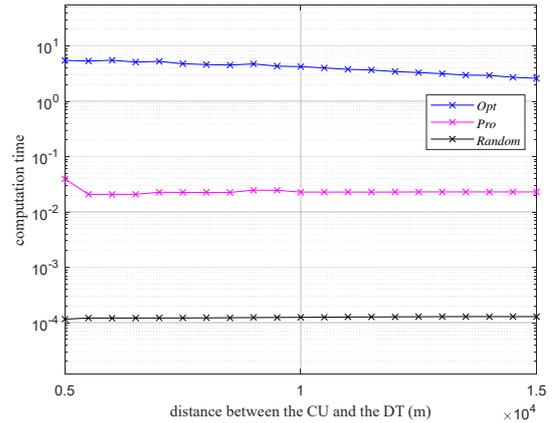

**FIGURE 9.** Computation time vs. the distance between the CU and the DT.

## V. CONCLUSIONS

This paper considers the energy-efficient wireless resource allocation issue for cooperative D2D communications. We developed a hybrid centralized-distributed scheme based on DRL and the KM algorithm to optimize joint spectrum allocation, power control, and link matching. Firstly, we took WSEE as the performance metric and formulated this joint optimization as an NP-hard MINLP problem. Secondly, we proposed a two-stage low-complexity tractable solution, which broke down the original joint problem into a weighted bipartite graph matching and a series of nonconvex spectrum allocation and power control subproblems between potentially matched cellular and D2D link pairs. Thirdly, we leveraged distributed DRL at DTs for each joint spectrum allocation and power control subproblem while guaranteeing the required QoS of each link. Fourthly, we determined the link matching at the BS via the KM algorithm in a centralized manner. At the same time, we discussed the approach to reduce signal exchange between a D2D link and its potential matched cellular links by employing the cooperative link set. Finally, we compared the proposed scheme with some heuristic schemes through simulations. The proposed scheme was demonstrated to achieve near-optimal performance and significantly enhance the network convergence speed with low signaling overheads.